\documentclass[pra,aps,showpacs,nofootinbib,reprint]{revtex4-1}
\usepackage[dvips]{graphicx} 
\usepackage{natbib}
\usepackage{color}
\begin{document}
\title{Cosmic Microwave Background Bispectrum from the
Lensing--Rees-Sciama Correlation Reexamined: Effects of Non-linear
Matter Clustering}
\author{Veronika Junk}
\affiliation{University Observatory Munich, 
  Scheinerstr. 1, D-81679 Munich, Germany}
\author{Eiichiro Komatsu}
\affiliation{Texas Cosmology Center and the Department of Astronomy,
The University of Texas at Austin, 1 University Station, C1400, Austin,
TX 78712}
\affiliation{Kavli Institute for the Physics and Mathematics of the
Universe, Todai Institutes for Advanced Study, the University of Tokyo,
Kashiwa, Japan 277-8583 (Kavli IPMU, WPI)}
\affiliation{Max Planck Institut f\"ur Astrophysik, Karl-Schwarzschild-Str. 1, 85741 Garching, Germany}
\begin{abstract}
 The bispectrum of the cosmic microwave background (CMB) generated by a
 correlation between a time-dependent gravitational potential and the
 weak gravitational lensing effect provides a direct measurement of the
 influence of dark energy on CMB. This bispectrum is also known to yield
 the most 
 important contamination of the so-called ``local-form'' primordial
 bispectrum, which can be used to 
 rule out all single-field inflation models. In this paper, we reexamine
 the effect of non-linear matter clustering on this bispectrum. We
 compare three different approaches: the 3rd-order perturbation theory
 (3PT), and two empirical fitting formulae available in the
 literature, finding that 
 detailed modeling of non-linearity appears to be not very
 important, as most of the signal-to-noise comes from the squeezed
 triangle, for which the correlation in the linear regime
 dominates. The expected signal-to-noise ratio for an
 experiment dominated by the cosmic variance up to $l_{\rm
 max}=1500$ is about 5, which is much smaller than the previous
 estimates including non-linearity, but agrees with the estimates based
 on the linear calculation. We 
 find that the difference between the linear and non-linear predictions
 is undetectable, and does not alter the contamination of the
 local-form primordial non-Gaussianity.
\end{abstract} 
\maketitle
%%%%%%%%%%%%%%%%%%%%%%%%%%%%%%%%%%%%%%%%%%%%%%%%%%%%%%%%%%%%%%%%%%%%%%%%%%%%%%%%
%%%%%%%%%%%%%%%%%%%%%%%%%%%%%%%%%%%%%%%%%%%%%%%%%%%%%%%%%%%%%%%%%%%%%%%%%%%%%%%%
\section{Introduction}
%%%%%%%%%%%%%%%%%%%%%%%%%%%%%%%%%%%%%%%%%%%%%%%%%%%%%%%%%%%%%%%%%%%%%%%%%%%%%%%%
%%%%%%%%%%%%%%%%%%%%%%%%%%%%%%%%%%%%%%%%%%%%%%%%%%%%%%%%%%%%%%%%%%%%%%%%%%%%%%%%
A time-dependent gravitational potential changes the temperature of the
cosmic microwave background (CMB) as $\delta T/T=2\int dt~(\partial
\Psi/\partial t)$ \cite{Sachs_Wolfe_1967,Rees_Sciama_1968}, where $\Psi$ is a
perturbation to the time-time component of the
Friedmann-Lema\^itre-Robertson-Walker metric. As $\Psi$ is constant
during the matter-dominated era, a detection of this effect directly
shows that the universe is not completely matter-dominated, but has
contributions from either spatial curvature or dark energy. Given the
tight constraint on the spatial curvature we have from the current
cosmological data \cite{Komatsu_2011}, a 
detection of this effect is considered as the direct evidence for the
effect of dark energy on the growth of structure \cite{Crittenden_Turok_1995}.

The weak gravitational lensing effect caused by matter density
fluctuations between us and the last scattering surface also changes the
temperature of CMB, by shifting the observed directions of photons as
$T(\hat{{\mathbf n}})\to T(\hat{{\mathbf n}}+{\mathbf
d})=T(\hat{{\mathbf n}})+{\mathbf d}\cdot \nabla T(\hat{{\mathbf
n}})+\dots$ \cite{Lewis_Challinor_2006}. Here, the
deflection angle ${\mathbf d}$ is given by ${\mathbf d}=2\int dr
\frac{r_*-r}{rr_*}\nabla\Psi(\hat{\mathbf n}r)$. As the same $\Psi$
enters in both effects, these two effects are correlated, yielding
a non-zero 3-point correlation (bispectrum) in the CMB
\cite{Goldberg_Spergel_1999}. Therefore, this bispectrum can be used to
probe the nature of dark energy \cite{Verde_Spergel_2002}. 

The time-dependence of $\Psi$ is caused by two effects: one is the
linear growth, and the other is the {\it non-linear} growth. The former
effect is caused by dark energy slowing down the growth of structure,
leading to a decay of $\Psi$. The latter effect is caused by non-linear
evolution of density fluctuations, leading to a growth of
$\Psi$. Following the literature, we shall call the former the
``integrated Sachs--Wolfe (ISW; \cite{Sachs_Wolfe_1967}) effect,'' and the latter the
``Rees--Sciama (RS; \cite{Rees_Sciama_1968}) effect.'' In this paper, we
shall calculate the 
bispectrum generated by the lensing-ISW correlation on large scales as
well as the lensing-RS correlation on small scales. 

The main focus of this paper is the lensing-RS correlation. This
correlation has been studied in the past
\cite{Verde_Spergel_2002,Baccigalupi_2003,*Giovi_2005,Mangilli_Verde_2009}
with different empirical methods for computing the non-linear matter power
spectrum. In this paper, we systematically compare two
empirical methods used in the literature and the 3rd-order perturbation
theory (3PT), which is applied to the lensing-RS bispectrum for the
first time in this paper. We find that, while these different methods
yield somewhat different results for the lensing-RS bispectrum, the
differences are too small to detect or affect our interpretation of the data.

In Sec.~\ref{LRS-CMB_bispectrum}, we review the lensing-RS
bispectrum. In Sec.~\ref{nonlinear_pk}, we compare three methods for
computing the non-linear matter power spectrum. In
Sec.~\ref{bispectrum_analysis}, we compare the lensing-RS
cross-correlation power spectrum, $Q(l)$, computed with different
non-linear matter power spectra. 
In Sec.~\ref{results}, we calculate the expected signal-to-noise ratio of the
lensing-RS bispectrum and the $\chi^2$ differences between the linear
model and various non-linear models. 
In Sec.~\ref{fNL}, we study the effects of non-linearity on the
contamination of the local-form primordial non-Gaussianity parameter,
$f_{\rm NL}$. 
We conclude in Sec~\ref{conclusions}.

Throughout this paper, we shall use the cosmological parameters given by
the WMAP 5-year best-fit parameters (WMAP+BAO+$H_0$ ML; \cite{Komatsu_2009}):
$\Omega_M=0.277$, $\Omega_\Lambda=0.723$, $h=0.702$, $n_s=0.962$, and
$\sigma_8=0.817$.
%%%%%%%%%%%%%%%%%%%%%%%%%%%%%%%%%%%%%%%%
%%%%%%%%%%%%%%%%%%%%%%%%%%%%%%%%%%%%%%%%
%%%%%%%%%%%%%%%%%%%%%%%%%%%%%%%%%%%%%%%%
\section{Lensing-RS Bispectrum}
\label{LRS-CMB_bispectrum}
%%%%%%%%%%%%%%%%%%%%%%%%%%%%%%%%%%%%%%%%
%%%%%%%%%%%%%%%%%%%%%%%%%%%%%%%%%%%%%%%%
%%%%%%%%%%%%%%%%%%%%%%%%%%%%%%%%%%%%%%%%
Let us use  spherical harmonics to expand the observed temperature anisotropy,
$\delta T(\hat{\mathbf n})/T=\sum a_{lm}Y_{lm}(\hat{\mathbf n})$, as
well as the ``lensing potential,'' $\Theta(\hat{\mathbf
n})=\sum_{lm}\Theta_{lm}Y_{lm}(\hat{\mathbf n})$, defined by ${\mathbf
d}\equiv \nabla \Theta$. Then, the CMB bispectrum generated by the
lensing-RS correlation is given by 
\citep{Goldberg_Spergel_1999,Komatsu_Spergel_2001}  
\begin{eqnarray}
\nonumber
B^{m_1 m_2 m_3}_{l_1 l_2 l_3} &\equiv& \langle
a_{l_1m_1}a_{l_2m_2}a_{l_3m_3}\rangle \\
&=& {\cal G}^{m_1 m_2 m_3}_{l_1 l_2 l_3}
\left[\frac{l_1(l_1+1)-l_2(l_2+1)+l_3(l_3+1)}{2} \right.
\nonumber \\
&\times&
\left. C^{P}_{l_1} \langle \Theta^{*}_{l_3 m_3} a^{\mathrm{ISW}}_{l_3
m_3} \rangle + 5 \, perm.\right].
\label{RS_bispectrum}
\end{eqnarray}
Here, $C_l^{P}$ is the primary CMB power spectrum without
lensing, and $a_{lm}^{\rm ISW}$ are the spherical harmonics coefficients
of the ISW (or RS) effect. 

However, Eq.~(\ref{RS_bispectrum}) is the leading-order contribution
which is accurate only to $\sim 10$\% level at $l\gtrsim 2000$. Lewis,
Challinor and Hanson \cite{Lewis_2011} have shown that the
sub-leading-order correction can be incorporated by simply replacing
$C_l^P$ above with the {\it lensed} CMB power spectrum, $C_l$:
\begin{eqnarray}
\nonumber
B^{m_1 m_2 m_3}_{l_1 l_2 l_3} &=& {\cal G}^{m_1 m_2 m_3}_{l_1 l_2 l_3}
\left[\frac{l_1(l_1+1)-l_2(l_2+1)+l_3(l_3+1)}{2} \right.
\nonumber \\
&\times&
\left. C_{l_1} \langle \Theta^{*}_{l_3 m_3} a^{\mathrm{ISW}}_{l_3
m_3} \rangle + 5 \, perm.\right].
\label{RS_bispectrum2}
\end{eqnarray}
We shall use this formula for computing the lensing-ISW (RS) bispectrum. Note that this prescription of replacing $C_l^P$ with $C_l$ is accurate only for the bispectrum in the squeezed configuration where one of the wavenumbers, say, $l_1$, is much smaller than the other two, i.e., $l_1\ll l_2\approx l_3$. This is sufficient for our purpose because the signal-to-noise ratio of the lensing-ISW (or RS) bispectrum is dominated by the squeezed configuration.

The Gaunt integral, ${\cal
G}^{m_1m_2m_3}_{l_1 l_2 l_3}$,  is defined as
\begin{eqnarray}
\nonumber
{\cal G}^{m_1 m_2 m_3}_{l_1 l_2 l_3} &\equiv&
\sqrt{\frac{(2l_1+1)(2l_2+1)(2l_3+1)}{4 \pi}}\\
&\times& \left(
\begin{array} {ccc} 
\l_1 & \l_2 & \l_3 \\
       0     &  0     &  0   
\end{array}
\right)
\left(
\begin{array}{ccc} 
 l_1 & l_2 & l_3 \\
 m_1 & m_2 & m_3
\end{array}
\right).
\label{Gaunt_integral}
\end{eqnarray}
Assuming statistical isotropy of the universe, rotational invariance
implies that one can average over orientation of triangles (i.e., $m$'s)
to obtain the angle-averaged bispectrum \cite{Spergel_Goldberg_1999}:
\begin{eqnarray}
\nonumber
B_{l_1l_2l_3} &\equiv& \sum_{m_1m_2m_3}\left(\begin{array}{ccc} 
 l_1 & l_2 & l_3 \\
 m_1 & m_2 & m_3
\end{array}
\right)B^{m_1 m_2 m_3}_{l_1 l_2 l_3} \\
\nonumber
&=&\sqrt{\frac{(2l_1+1)(2l_2+1)(2l_3+1)}{4 \pi}}
\left(
\begin{array}{ccc} 
l_1 & l_2 & l_3 \\
0 & 0 & 0 
\end{array}
\right)\\
\nonumber
&\times&\left[
\frac{
l_1(l_1+1)-l_2(l_2+1)+l_3(l_3+1)
}{2}\right.\\
& &\times\left.
C_{l_1} \langle \Theta^{*}_{l_3m_3} a^{\mathrm{ISW}}_{l_3m_3} \rangle + 5 \, perm.\right].
\end{eqnarray}
Here, the cross-power spectrum of the lensing potential and the ISW (or
RS) effect, $Q(l)\equiv \langle \Theta^{*}_{l_3m_3}
a^{\mathrm{ISW}}_{l_3m_3} \rangle$, is given by \cite{Goldberg_Spergel_1999,Verde_Spergel_2002}
\begin{eqnarray}
\nonumber
Q(l)    &\equiv& \langle \Theta^{*}_{l_3m_3}
a^{\mathrm{ISW}}_{l_3m_3} \rangle\\
&=&
2 \int^{z_{*}}_{0}dz{\frac{r(z_{*})-r(z)}{r(z_{*})r(z)^3}
    \left.\frac{\partial P_{\Psi}(k,z)}{\partial z}\right|_{k =
    l/r(z)}},
\label{Ql3_amplitude_bispectrum}
\end{eqnarray} 
where $z_{*}=1090$ is the redshift of the last scattering surface, 
$P_{\Psi}$ the power spectrum of the Newtonian potential: 
\begin{equation}
\label{P_phi}
P_{\Psi}(k,z) =  
\left(
\frac{3}{2} \Omega_M
\right)^2 
\left(
\frac{H_0}{k}
\right)^4 P(k,z) (1 + z)^2,
\end{equation}
and $P(k)$ the power spectrum of matter density fluctuations,
$\delta_M$. This result follows from the Poisson equation (in natural
units): $k^2\Psi(k,z)=-4\pi G\rho_M(z)a^2(z)\delta_M(k,z)=-4\pi
G\rho_{M0}\delta_M(k,z)(1+z)= -\frac32\Omega_MH_0^2\delta_M(k,z)(1+z)$.

On large scales where the scale-invariant spectrum, $P_\Psi\propto
1/k^3$, is still preserved, the cross-power spectrum goes as
$Q(l)\propto 1/l^3$. On the other hand, the primary power spectrum goes
as $C_l^{\rm P}\propto 1/l^2$. On smaller scales, $Q(l)$ falls even
faster than $1/l^3$. This implies that the bispectrum peaks at the
``squeezed triangle,'' for which one of $l$'s is much smaller than the
other two (e.g., $l_3\ll l_1\simeq l_2$ if we order multipoles such that
$l_3\le l_2\le l_1$), and the smallest $l$ corresponds to $l$ of
$Q(l)$. This observation suggests that the signal would be dominated by
$Q(l)$ in the small $l$ for which matter fluctuations can still be
treated as linear perturbations, and thus the detailed modeling of
non-linear fluctuations may not be necessary. We will confirm this
observation in this paper. 

The remaining task is to calculate $P(k,z)$, including non-linear matter
clustering. 
%%%%%%%%%%%%%%%%%%%%%%%%%%%%%%%%%%%%%%%%
%%%%%%%%%%%%%%%%%%%%%%%%%%%%%%%%%%%%%%%%
%%%%%%%%%%%%%%%%%%%%%%%%%%%%%%%%%%%%%%%%
\section{Nonlinear matter power spectrum}
\label{nonlinear_pk}
%%%%%%%%%%%%%%%%%%%%%%%%%%%%%%%%%%%%%%%%
%%%%%%%%%%%%%%%%%%%%%%%%%%%%%%%%%%%%%%%%
%%%%%%%%%%%%%%%%%%%%%%%%%%%%%%%%%%%%%%%%
%%%%%%%%%%%%%%%%%%%%%%%%%%%%%%%%%%%%%%%%%%%%%%%%%%%%%%%%%%%%%%%%%%%%%%%%%%%%%%%%%%%%%%%%%%%%%%%%%%%%%%%%%%%%%
%%%%%%%%%%%%%%%%%%%%%%%%%%%%%%%%%%%%%%%%%%%%%%%%%%%%%%%%%%%%%%%%%%%%%%%%%%%%%%%%%%%%%%%%%%%%%%%%%%%%%%%%%%%%%
%%%%%%%%%%%%%%%%%%%%%%%%%%%%%%%%%%%%%%%%%%%%%%%%%%%%%%%%%%%%%%%%%%%%%%%%%%%%%%%%%%%%%%%%%%%%%%%%%%%%%%%%%%%%%
\subsection{3rd-order Perturbation Theory (3PT)}
\label{nonlinear_descriptions}
%%%%%%%%%%%%%%%%%%%%%%%%%%%%%%%%%%%%%%%%%%%%%%%%%%%%%%%%%%%%%%%%%%%%%%%%%%%%%%%%%%%%%%%%%%%%%%%%%%%%%%%%%%%%%
%%%%%%%%%%%%%%%%%%%%%%%%%%%%%%%%%%%%%%%%%%%%%%%%%%%%%%%%%%%%%%%%%%%%%%%%%%%%%%%%%%%%%%%%%%%%%%%%%%%%%%%%%%%%%
%%%%%%%%%%%%%%%%%%%%%%%%%%%%%%%%%%%%%%%%%%%%%%%%%%%%%%%%%%%%%%%%%%%%%%%%%%%%%%%%%%%%%%%%%%%%%%%%%%%%%%%%%%%%%
Higher-order perturbation theory is a promising approach for computing
non-linear evolution of matter density fluctuations
\cite{Bernardeau_2002}. This is especially true at high redshifts
($z>1$), where non-linearity is not too strong
\cite{Jeong_Komatsu_2006}. The lensing-RS correlation has been studied
using the 3rd-order perturbation theory (3PT) by \cite{Nishizawa_2008},
who found a reasonable agreement between the 3PT prediction and the data
obtained from the $N$-body simulation.\footnote{Due to a page
limitation, Ref.~\cite{Nishizawa_2008} did not report on the details of the
3PT results in the published version; however, the details are reported
in arXiv:0711.1696.}

The matter power spectrum including the next-to-leading order non-linear
correction is given by \cite{Vishniac_1983,*Fry_1984,*Goroff_Grinstein_Rey_1986,*Suto_Sasaki_1991,*Makino_Sasaki_Suto_1992,*Jain_Bertschinger_1994}
\begin{equation}
\label{powerspectrum_PT}
P(k,z) = [D(z)]^2 P_{11}(k) + [D(z)]^4[2 P_{13}(k)+P_{22}(k)],
\end{equation}
where $D(z)$ is a suitably normalized linear growth factor (which
is proportional to the scale factor during the matter era), $P_{11}(k)$
is the linear power spectrum at an arbitrary initial time, $z_i$, at
which $D(z_i)$ is normalized to unity, and $P_{22}(k)$ and $P_{13}(k)$ are
given by
\begin{equation}
P_{22}(k) = 2 \int \frac{d^3 q}{(2 \pi)^3} P_{11}(q)
P_{11}(|{\bf{k}}-{\bf{q}}|)\left[
F^{(s)}_2({\bf{q}},{\bf{k}}-{\bf{q}})
\right]^2, 
\label{PT_pk22}
\end{equation}
where
\begin{equation}
F^{(s)}_2({\bf{k}}_1,{\bf{k}}_2) = \frac{5}{7} + \frac{2}{7} \frac{({\bf{k}}_1 \cdot {\bf{k}}_2)^2}{k^2_1 k^2_2} + 
\frac{{\bf{k}}_1 \cdot {\bf{k}}_2}{2}
\left(
\frac{1}{k^2_1} + \frac{1}{k^2_2}
\right), 
\end{equation}
and
\begin{eqnarray}
\nonumber
2 P_{13}(k) &=& \frac{2 \pi k^2}{252} P_{11}(k) \int^{\infty}_{0}
\frac{dq}{(2 \pi)^3} P_{11}(q) \\
\nonumber
&\times &
\left[
100 \frac{q^2}{k^2} -158 + 12 \frac{k^2}{q^2} - 42 \frac{q^4}{k^4}\right.\\
\label{P13_eq}
&+ & \left.\frac{3}{k^5 q^3} (q^2-k^2)^3 (2 k^2 + 7 q^2) 
\ln{
\left(
\frac{k+q}{|k-q|}
\right)
}
\right].
\end{eqnarray}

The 3PT is an attractive approach, as it provides the {\it exact} calculation
in the quasi linear regime where the perturbative expansion is still
valid. This should be contrasted with the empirical approaches described
below: they are calibrated using numerical simulations with a specific
set of cosmological parameters, and thus cannot be easily extended to
other cosmological models, such as dynamical dark energy models. 

A disadvantage of the 3PT is that its validity is limited to the quasi
linear regime, and thus the result on very small scales cannot be
trusted. One can check the validity of the 3PT calculation by comparing
it to a direct numerical simulation
\cite{Jeong_Komatsu_2006,Nishizawa_2008}; or, one can 
compare an empirical formula calibrated to a specific cosmological
model, to the 3PT calculation using the same cosmological model. We
shall adopt the latter approach in this paper.
%%%%%%%%%%%%%%%%%%%%%%%%%%%%%%%%%%%%%%%%%%%%%%%%%%%%%%%%
\subsection{Empirical Models} 
Empirical approaches, which are calibrated using $N$-body simulations,
have an advantage that they can, in principle, describe the matter power
spectrum in a highly non-linear regime where the perturbative expansion
breaks down. We shall use one of the popular methods, called the ``halo
model,'' \cite{Seth_Cooray_2002} for checking the validity of the 3PT
for computing the 
lensing-RS power spectrum. We then compare these results with another
empirical model \citep{Ma_1999} used by most of the previous work on the
lensing-RS bispectrum.
%%%%%%%%%%%%%%%%%%%%%%%%%%%%%%%%%%%%%%%%%%%%%%%%%%%%%%%%
\subsubsection{Halo Model (HALOFIT)}
%%%%%%%%%%%%%%%%%%%%%%%%%%%%%%%%%%%%%%%%%%%%%%%%%%%%%%%%
In the halo model, the matter power spectrum is decomposed into two
pieces: one that arises from two-point correlations between dark matter
particles residing in two different dark matter halos (2-halo term), and
another that arises from two-point correlations between dark matter
particles residing in a single dark matter halo (1-halo term). The
former contribution 
is given approximately by the linear matter power spectrum, whereas the
latter contribution is given by the density profile of dark matter
halos. This splitting between the 2- and 1-halo terms is somewhat
artificial, and thus the halo model approach would not provide an
accurate description of the non-linear matter power spectrum, unless it
is calibrated by numerical simulations.

One popular calibrated formula is due to Smith et
al.~\cite{Smith_Peacock_Jenkins_2003}, which will be called ``HALOFIT.''
They model the power spectrum as 
$P(k)=P_Q(k)+P_H(k)$, where $P_Q(k)$ is the quasi-linear 2-halo term: 
\begin{equation}
 P_Q(k)=P_{11}(k)\frac{\left[1+k^3P_{11}(k)/(2\pi^2)\right]^{\beta_n}}{1+\alpha_n
  k^3P_{11}(k)/(2\pi^2)}\exp\left[-f(y)\right],
\end{equation}
where $f(y)\equiv y/4+y^2/8$ with $y=k/k_\sigma$, and $\alpha_n$,
$\beta_n$, and $k_\sigma$ 
are free parameters which need to be determined from simulations
(see Appendix C of \cite{Smith_Peacock_Jenkins_2003}). The second term,
$P_H(k)$, is the 1-halo term:
\begin{equation}
 P_H(k)=\frac1{1+\mu_n/y+\nu_n/y^2}\frac{a_ny^{3f_1}}{1+b_ny^{f_2}+[c_nf_3y]^{3-\gamma_n}},
\end{equation}
where $\mu_n$, $\nu_n$, $a_n$, $b_n$, $c_n$, $\gamma_n$, $f_1$, $f_2$, and
$f_3$ are free parameters which need to be determined from
simulations. As one of the cosmological models for which these functions 
are calibrated is a $\Lambda$CDM model with $\Omega_M=0.3$,
$\Omega_\Lambda=0.7$, $h=0.7$, and $\sigma_8=0.9$, which is close to the
parameters we adopt in this paper, this model can be used to check the
validity of the 3PT results in the non-linear regime. However, we remind
the readers that HALOFIT is not guaranteed to provide accurate results
for the cosmological models that are not explored in
Ref.~\cite{Smith_Peacock_Jenkins_2003}, such as dynamical dark energy
or massive neutrino models. Ref.~\cite{Mangilli_Verde_2009} also used
HALOFIT for computing the lensing-RS bispectrum.

\subsubsection{HKLM Scaling Model (MA99)}
\label{sec:ma99}
Another empirical formula is based on the idea originally put forward by HKLM
\cite{Hamilton_1991} for the real-space two-point correlation
function. This idea has been applied to the power spectrum by
\cite{Peacock_Dodds_1994,Peacock_Dodds_1996}. Then, Ma et
al. \cite{Ma_1999} have extended the calibration to include models with
dynamical dark energy. We shall use the formula of Ma et al., and call
it ``MA99.''

HKLM postulates that the non-linear correlation function is a universal
function of the linear correlation function, once the length scale (or
the wavenumber) is
rescaled by the mass conservation (i.e., transformation from Lagrangian
to Eulerian coordinates). The form of this universal function needs to
be found from numerical simulations. Ma et al. \cite{Ma_1999} find 
\begin{equation}
 \label{powerspectrum_Ma}
\frac{k^3P(k,z)}{2\pi^2} = 
G 
\left[
\frac{k^3_0P_{11}(k_0,z)/(2\pi^2)}{g^{3/2}_{0}[\sigma_8(z)]^{\beta}} 
\right]\frac{k^3_0P_{11}(k_0,z)}{2\pi^2},
\end{equation}
where $\beta=0.83$, $\sigma_8(z)$ is related to the present-day $\sigma_8$ by 
 $\sigma_8(z)=[D(z)/D(0)]\sigma_8$, and $g_0$ is defined by
$g_0\equiv |w|^{1.3|w|-0.76} g(0)$.
Here, $g(z)\propto (1+z)D(z)$, but it is normalized such that $g(z_i)=1$
during the matter-dominated era (e.g., $z_i=30$).\footnote{For example,
$g(0)=0.7646$ for $\Omega_M=0.277$, $\Omega_\Lambda=0.723$ and $w=-1$.}
The Lagrangian wavenumber, $k_0$, is related to the Eulerian wavenumber,
$k$, as
\begin{equation}
\label{scaletrans}
 k_0\equiv \frac{k}{[1+k^3P(k,z)/(2\pi^2)]^{1/3}}.
\end{equation}
The function, $G(x)$, is given by
\begin{equation}
\label{function_G}
G(x) = 
\left[ 
1+ \ln{(1+0.5 x)} 
\right]
\frac{1 + 0.02 x^4 + c_1 x^8/g(z)^3}{1 + c_2 x^{7.5}}, 
\end{equation}
with $c_1 = 1.08\times 10^{-4}$ and $c_2=2.10 \times 10^{-5}$.

How do we compute $P(k,z)$ using this formula?\footnote{While this
formula has been used by most of the previous work on the
lensing-RS bispectrum
\cite{Verde_Spergel_2002,Baccigalupi_2003,*Giovi_2005},
in all cases it has been implemented incorrectly. In the previous work,
the authors used Eq.~(\ref{powerspectrum_Ma}) with $k_0$ in the argument
of $G(x)$ replaced by $k$. This is not the implementation proposed by
the original paper \cite{Ma_1999}, and goes against the original
proposal made by HKLM. This observation gave an initial motivation for
our work.} 
\begin{itemize}
 \item[1.] Compute the linear power spectrum, $P_{11}(k_0,z)$, for a
	   given Lagrangian wavenumber, $k_0$.
 \item[2.] Compute $G(x)$ in Eq.~(\ref{powerspectrum_Ma}), and
	   multiply it by $k_0^3P_{11}(k_0,z)/(2\pi^2)$ to obtain
	   $k^3P(k,z)/(2\pi^2)$. 
 \item[3.] Compute the Eulerian wavenumber, $k$, using
	   Eq.~(\ref{scaletrans}).
 \item[4.] Compute $P(k,z)$ from $k^3P(k,z)/(2\pi^2)$ times $(2\pi^2)/k^3$.
\end{itemize}

%%%%%%%%%%%%%%%%%%%%%%%%%%%%%%%%%%%%%%%%%%%%%%%%%%%%%%%%%%%%%%%%%%%%%%%%
%%%%%%%%%%%%%%%%%%%%%%%%%%%%%%%%%%%%%%%%%%%%%%%%%%%%%%%%%%%%%%%%%%%%%%%%
\subsection{Comparing  $P(k,z)$}
\label{comparison_PT_Ma}
\begin{figure}[t]
\begin{center}
\includegraphics[width=0.49\textwidth]{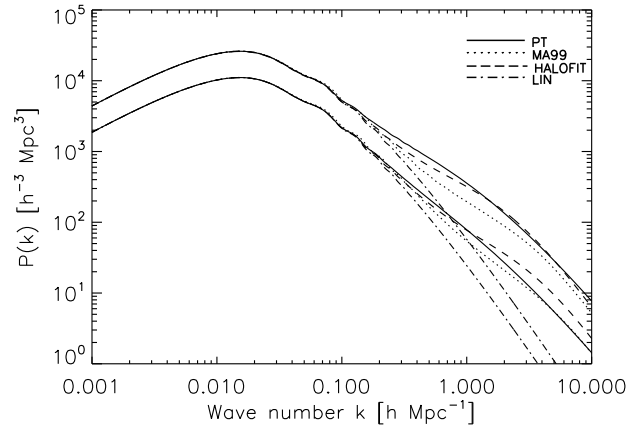}
\caption[]
{
 Comparison of non-linear power spectra computed from 3PT (solid
 line), MA99 (dotted line), and HALOFIT (dashed line). The
 linear power spectrum is shown by the dashed-dotted line. The upper and
 lower curves show $P(k,z)$ for $z=0.1$ and 1, respectively.
}
\label{Power_compare_PT_Ma}
\end{center}
\end{figure}

%%%%%%%%%%%%%%%%%%%%%%%%%%%%%%%%%%%%%%%%%%%%%%%%%%%%%%%%%%%%%%%%%%%%%%%%%%%%%%%%%%%%%%%%%%%%%%%%%%%%%%%%%%%%%
%%%%%%%%%%%%%%%%%%%%%%%%%%%%%%%%%%%%%%%%%%%%%%%%%%%%%%%%%%%%%%%%%%%%%%%%%%%%%%%%%%%%%%%%%%%%%%%%%%%%%%%%%%%%%
In Figure~\ref{Power_compare_PT_Ma}, we show non-linear power spectra at
$z=0.1$ and 1 computed from  3PT (solid line), MA99 (dotted
line), and HALOFIT (dashed line). We find that, somewhat surprisingly,
HALOFIT and Ma et al. for $z=0.1$ are fairly discrepant at $k\approx
0.3-3~h~{\rm Mpc}^{-1}$. (MA99 underestimates the power relative to
HALOFIT.) 

In order to identify the origin of this discrepancy, we have
also compared MA99 and HALOFIT with the formula by Peacock and Dodds
(1996; PD96) \cite{Peacock_Dodds_1996} (not shown in Figure~1; in order to use
their formula, it is necessary to use the smooth linear power spectrum
without baryonic oscillations; thus, we used the smooth power spectrum
given in \cite{Eisenstein_Hu_1998}). We find that PD96 and HALOFIT agree
well, which is consistent with the finding of
\cite{Mangilli_Verde_2009}. However, MA99 and PD96, which are based on
the same HKLM idea, differ significantly. This probably indicates that
the difference already existed at the level of N-body simulations
used by MA99 and PD96. Given that the latest HALOFIT formula has been
shown to provide excellent fits to a wide range of N-body simulations,
we conclude that PD96 and HALOFIT are more accurate than MA99. This
suggests that the previous work based on MA99
\cite{Verde_Spergel_2002,Baccigalupi_2003,*Giovi_2005} would require a
reexamination. 

%%%%%%%%%%%%%%%%%%%%%%%%%%%%%%%%%%%%%%%%%%%%%%%%%%%%%%%%%%%%%%%%%%%%%%%%%%%%%%%%%%%%%%%%%%%%%%%%%%%%%%%%%%%%%
%%%%%%%%%%%%%%%%%%%%%%%%%%%%%%%%%%%%%%%%%%%%%%%%%%%%%%%%%%%%%%%%%%

On the other hand, the 3PT results are close to HALOFIT, but lie
slightly above it at $k\approx
0.2-1~h~{\rm Mpc}^{-1}$. This is a known result: at a low redshift, the
3PT tends to overpredict the non-linear power spectrum
\cite{Jeong_Komatsu_2006}. While this is an issue for the 1\%-level precision
cosmology using the galaxy power spectrum, the discrepancy at this level
may not be so bad for the calculation of the lensing-RS correlation, as
the statistical error on the expected total signal-to-noise of the
measurement of 
the lensing-RS bispectrum is modest ($S/N\lesssim 10$). We thus take this as
an encouraging sign and move on. 

%%%%%%%%%%%%%%%%%%%%%%%%%%%%%%%%%%%%%%%%%%%%%%%%%%%%%%%%%%%%%%%%%%%%%%%%%%%%%%%%%%%%%%%%%%%%%%%%%%%%%%%%%%%%%
%%%%%%%%%%%%%%%%%%%%%%%%%%%%%%%%%%%%%%%%%%%%%%%%%%%%%%%%%%%%%%%%%%%%%%%%%%%%%%%%%%%%%%%%%%%%%%%%%%%%%%%%%%%%%
%%%%%%%%%%%%%%%%%%%%%%%%%%%%%%%%%%%%%%%%%%%%%%%%%%%%%%%%%%%%%%%%%%%%%%%%%%%%%%%%%%%%%%%%%%%%%%%%%%%%%%%%%%%%%
\section{Lensing-RS Cross-power Spectrum}
\label{bispectrum_analysis}
%%%%%%%%%%%%%%%%%%%%%%%%%%%%%%%%%%%%%%%%%%%%%%%%%%%%%%%%%%%%%%%%%%%%%%%%%%%%%%%%%%%%%%%%%%%%%%%%%%%%%%%%%%%%%
%%%%%%%%%%%%%%%%%%%%%%%%%%%%%%%%%%%%%%%%%%%%%%%%%%%%%%%%%%%%%%%%%%%%%%%%%%%%%%%%%%%%%%%%%%%%%%%%%%%%%%%%%%%%%
%%%%%%%%%%%%%%%%%%%%%%%%%%%%%%%%%%%%%%%%%%%%%%%%%%%%%%%%%%%%%%%%%%%%%%%%%%%%%%%%%%%%%%%%%%%%%%%%%%%%%%%%%%%%%
\subsection{$\partial \ln P_\Psi(k,z)/\partial z$}
\label{sec:derivative}
The essential ingredient of the lensing-RS bispectrum is the lensing-RS
cross-power spectrum, $Q(l)$, defined by
Eq.~(\ref{Ql3_amplitude_bispectrum}). In order to compute $Q(l)$, we
need derivatives of the potential power spectrum, $P_\Psi$, with respect
to redshifts, $\partial P_\Psi(k,z)/\partial z$. This is related to
derivatives of the density power spectrum, $P(k,z)$, as (see Eq.~(\ref{P_phi}))
\begin{equation}
\label{derivative_p_phi}
\frac{\partial\ln P_{\Psi}(k,z)}{\partial z}  = \frac{\partial \ln P(k,z)}{\partial z} + \frac{2}{(1+z)}.
\end{equation}
As $P(k,z)\propto (1+z)^{-2}$ for the linear matter power spectrum
during the matter-dominated era, $\partial P_{\Psi}(k,z)/\partial z$
vanishes for this case, as expected. When the universe is dominated by
curvature or dark energy, the first term is still negative but becomes
smaller than the second term, yielding $\partial P_{\Psi}(k,z)/\partial
z>0$. On the other hand, the 3PT result
(Eq.~(\ref{powerspectrum_PT})) shows that non-linear evolution gives a
term in $P(k,z)$ which goes as $(1+z)^{-4}$, and thus one obtains
non-zero  $\partial P_{\Psi}(k,z)/\partial z$ even during the
matter-dominated era. The sign is opposite: $\partial
P_{\Psi}(k,z)/\partial z<0$. 

%%%%%%%%%%%%%%%%%%%%%%%%%%%%%%%%%%%%%%%%%%%%%%%%%%%%%%%%%%%%%%%%%% 
\begin{figure}[t]
\begin{center}
\includegraphics[width=0.49\textwidth]{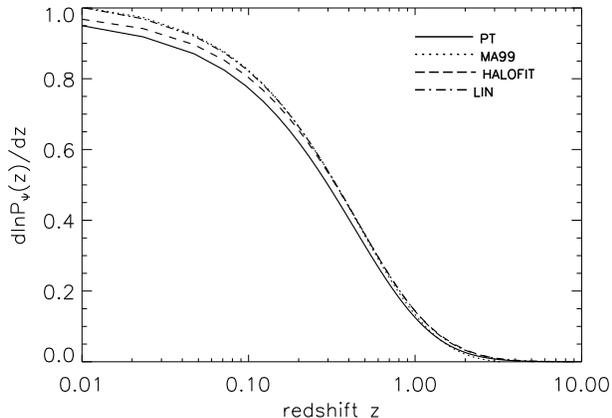}
\caption[]
{
 Comparison of $\partial \ln P_{\Psi}(k,z)/\partial z$ for
 $k=0.1~h\,\mathrm{Mpc}^{-1}$ as a function of
 $z$, computed from 3PT (solid 
 line), MA99 (dotted line), and HALOFIT (dashed line). 
}
\label{Pderiv_compare_PT_Ma_Giovi}
\end{center}
\end{figure}
%%%%%%%%%%%%%%%%%%%%%%%%%%%%%%%%%%%%%%%%%%%%%%%%%%%%%%%%%%%%%%%%%% 

We find that one needs to be quite careful about numerical accuracy when
computing $\partial \ln P_{\Psi}(k,z)/\partial z$. A stable result can be
obtained by the following method: compute $\ln P_\Psi(k,z)$ for
various redshifts separated by $\delta z=10^{-2}$, and then use a cubic spline
interpolation to evaluate the derivative. For $y(z)\equiv \ln P_{\Psi}(k,z)$, 
\begin{eqnarray}
\nonumber
 y'(z)&=&\frac{y(z_{\rm hi})-y(z_{\rm low})}{z_{\rm hi}-z_{\rm low}}\\
\nonumber
&+&\frac16[(3B^2-1)y''(z_{\rm hi})-(3A^2-1)y''(z_{\rm low})](z_{\rm hi}-z_{\rm low}),\\
\end{eqnarray}
where $A\equiv (z_{\rm hi}-z)/(z_{\rm hi}-z_{\rm low})$ and $B\equiv
(z-z_{\rm low})/(z_{\rm hi}-z_{\rm low})$, and $z_{\rm hi}$ and $z_{\rm
low}$ denote the pre-computed values of redshifts that are closest to
the chosen value of $z$. See Sec.~3.3 of \cite{NR}.
This method gives a highly
accurate  $\partial P_{\Psi}(k,z)/\partial z$ compared to a simpler
numerical differentiation such as $y(z)=[y(z+\delta z/2)-y(z-\delta 
z/2)]/\delta z$ or $y(z)=[y(z+\delta z)-y(z)]/\delta z$. We have
verified this using the 3PT results: for 3PT, one can calculate the
derivative exactly by differentiating Eq.~(\ref{powerspectrum_PT}) with
respect to $z$:
\begin{eqnarray}
\nonumber
 \frac{\partial P(k,z)}{\partial z}&=&2D(z)\frac{d D}{d z}\left\{P_{11}(k)\right.\\
\left.
+2[D(z)]^2\left[2P_{13}(k)+P_{22}(k)\right]\right\}.
\label{eq:derexact}
\end{eqnarray}
We find that the derivative from Eq.~(\ref{eq:derexact}) and that from
the cubic spline interpolation agree precisely.\footnote{We
suspect that 
any differences between our results presented in this paper and those
presented in the literature
\cite{Verde_Spergel_2002,Baccigalupi_2003,*Giovi_2005,Mangilli_Verde_2009}
can be explained by either an incorrect implementation of MA99 or an
inaccurate computation of the derivative or both. A code for reproducing
our results is available on {\sf http://www.mpa-garching.mpg.de/\textasciitilde{}komatsu/CRL/}}

In Figure~\ref{Pderiv_compare_PT_Ma_Giovi}, we show $\partial \ln
P_{\Psi}(k,z)/\partial z$ for $k=0.1~h~{\rm Mpc}^{-1}$ as a function of
$z$, computed from  3PT (solid 
line), MA99 (dotted line), HALOFIT (dashed line), and the linear
spectrum (dot-dashed line). At this wavenumber, they roughly agree with
each other to within 5\% at $z\ge 
0.01$. Non-linear evolution of matter fluctuations makes $\partial \ln
P_{\Psi}(k,z)/\partial z$ slightly smaller than the linear prediction. At this
wavenumber, the predicted non-linearity is the largest for 3PT, followed
by HALOFIT and MA99, which is consistent with
Figure~\ref{Power_compare_PT_Ma}.

\begin{figure}[t]
\begin{center}
\includegraphics[width=0.49\textwidth]{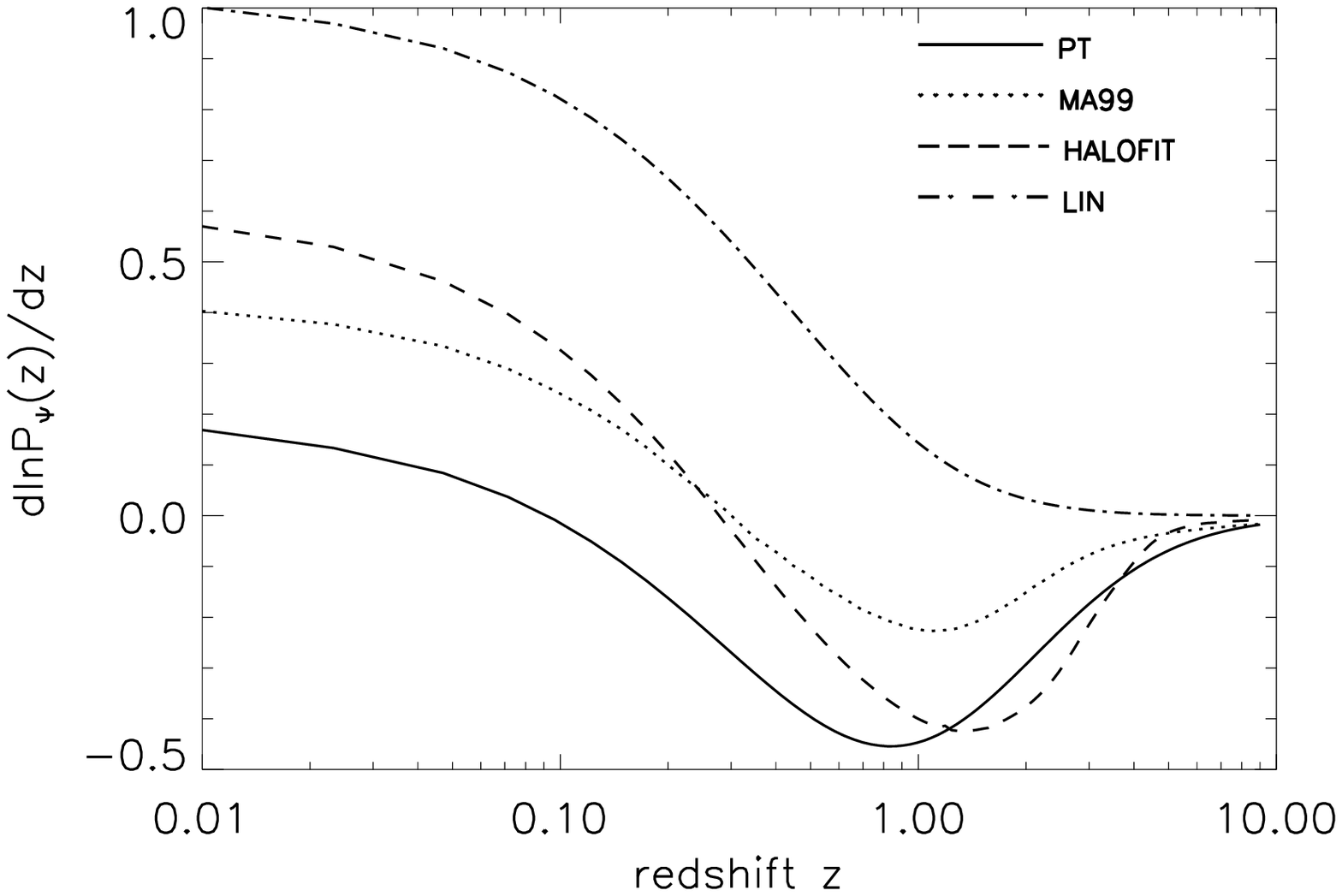}
\caption[]
{
 Same as Figure~\ref{Pderiv_compare_PT_Ma_Giovi}, but for
 $k=1~h\,\mathrm{Mpc}^{-1}$. 
}
\label{Pderiv_compare_PT_Ma_Giovi_highk}
\end{center}
\end{figure}

In Figure~\ref{Pderiv_compare_PT_Ma_Giovi_highk}, we show $\partial \ln
P_{\Psi}(k,z)/\partial z$ for $k=1~h~{\rm Mpc}^{-1}$. This regime is
quite non-linear, and thus we see a clear change in the sign of $\partial \ln
P_{\Psi}(k,z)/\partial z$ at a moderate redshift. (Recall that the
linear evolution due to dark energy gives a positive contribution to
$\partial \ln P_{\Psi}(k,z)/\partial z$, while the non-linear evolution
gives a negative contribution to $\partial \ln 
P_{\Psi}(k,z)/\partial z$.) 
The precise redshift at
which the sign changes depends on models of non-linearity: it is
$z\sim 0.1$ for 3PT while it is $z\sim 0.3$ for HALOFIT and MA99.

%%%%%%%%%%%%%%%%%%%%%%%%%%%%%%%%%%%%%%%%%%%%%%%%%%%%%%%%%%%%%%%%%% 
%%%%%%%%%%%%%%%%%%%%%%%%%%%%%%%%%%%%%%%%%%%%%%%%%%%%%%%%%%%%%%%%%% 

\subsection{$Q(l)$}
\begin{figure}[t]
\begin{center}
\includegraphics[width=0.49\textwidth]{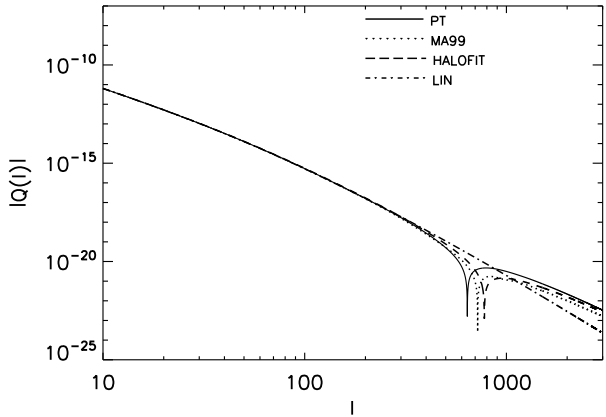}
\caption{Absolute values of the lensing-RS cross-power spectrum,
 $|Q(l)|$, as a function of multipoles, $l$, computed from 3PT (solid
 line), MA99 (dotted line), and HALOFIT (dashed line). The
 linear power spectrum result, which does not show any change of the sign,
 is shown by the dashed-dotted line. Note that the sign of $Q(l)$ is
 positive on large angular scales and negative on small angular scales.
}
\label{Ql}
\end{center}
\end{figure}

%%%%%%%%%%%%%%%%%%%%%%%%%%%%%%%%%%%%%%%%%%%%%%%%%%%%%%%%%%%%%%%%%%%%%%
With $\partial P_\Psi/\partial z$ computed, we now compute the
lensing-RS power spectrum, $Q(l)$, from
Eq.~(\ref{Ql3_amplitude_bispectrum}). In Figure~\ref{Ql}, we show
$|Q(l)|$ computed from 3PT (solid
 line), MA99 (dotted line) and HALOFIT (dashed line). The sign change
 due to non-linearity is seen, and the multipole at which the sign
 changes depends on models of non-linearity. It is $l\sim 640$, $700$,
 and $800$ for 3PT, MA99, and HALOFIT, respectively. 

%%%%%%%%%%%%%%%%%%%%%%%%%%%%%%%%%%%%%%%%%%%%%%%%%%%%%%%%%%%%%%%%%% 
%%%%%%%%%%%%%%%%%%%%%%%%%%%%%%%%%%%%%%%%%%%%%%%%%%%%%%%%%%%%%%%%%% 

%%%%%%%%%%%%%%%%%%%%%%%%%%%%%%%%%%%%%%%%%%%%%%%%%%%%%%%%%%%%%%%%%%
%%%%%%%%%%%%%%%%%%%%%%%%%%%%%%%%%%%%%%%%%%%%%%%%%%%%%%%%%%%%%%%%%%
%%%%%%%%%%%%%%%%%%%%%%%%%%%%%%%%%%%%%%%%%%%%%%%%%%%%%%%%%%%%%%%%%%
\section{Results}
\label{results}
\subsection{Signal-to-noise ratio}
%%%%%%%%%%%%%%%%%%%%%%%%%%%%%%%%%%%%%%%%%%%%%%%%%%%%%%%%%%%%%%%%%%
%%%%%%%%%%%%%%%%%%%%%%%%%%%%%%%%%%%%%%%%%%%%%%%%%%%%%%%%%%%%%%%%%%
%%%%%%%%%%%%%%%%%%%%%%%%%%%%%%%%%%%%%%%%%%%%%%%%%%%%%%%%%%%%%%%%%%

How well can we measure the lensing-RS bispectrum? 
The expected signal-to-noise ratio is given by \citep{Spergel_Goldberg_1999}
\begin{eqnarray}
\nonumber
\left(
\frac{S}{N}
\right)^2 &\equiv& 
\frac16
\sum_{2 \leq l_1 l_2 l_3\le l_{\mathrm{max}}}  
{\frac{B_{l_1 l_2 l_3}^2  }{C_{l_1}C_{l_2}C_{l_3}}}\\
&=&
\sum_{2 \leq l_1 \leq l_2 \leq l_3 < l_{\mathrm{max}}} 
{\frac{B_{l_1 l_2 l_3}^2  }{\Delta_{l_1l_2l_3}C_{l_1}C_{l_2}C_{l_3} }},
\label{Signal_to_Noisse_komat}
\end{eqnarray}
where $\Delta_{l_1l_2l_3}=1$ if all $l$'s are different,
$\Delta_{l_1l_2l_3}=2$ if two $l$'s are equal (isosceles configuration),
and $\Delta_{l_1l_2l_3}=6$ if all $l$'s are equal (equilateral
configuration). 

%%%%%%%%%%%%%%%%%%%%%%%%%%%%%%%%%%%%%%%%%%%%%%%%%%%%%%%%%%%%%%%%%% 
\begin{figure}[t]
\begin{center}
\includegraphics[width=0.49\textwidth]{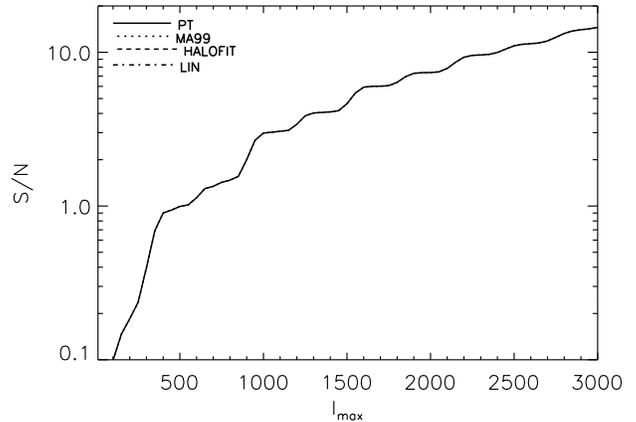}
\caption[]
{
 Expected signal-to-noise ratio of the lensing-RS bispectrum, $S/N$, as
 a function of the maximum multipole, $l_{\rm max}$. All non-linear
 models as well as the linear model give similar results. 
}
\label{sum_plots_PT_MA99_HALOFIT}
\end{center}
\end{figure}
%%%%%%%%%%%%%%%%%%%%%%%%%%%%%%%%%%%%%%%%%%%%%%%%%%%%%%%%%%%%%%%%%%%%

This formula assumes that non-Gaussianity is weak, and the covariance
matrix of the bispectrum can be approximated by the Gaussian piece,
$C_{l_1}C_{l_2}C_{l_3}$. However, Lewis, Challinor and Hanson
 have shown that there is a non-negligible contribution
from the non-Gaussian signal generated by the lensing-ISW bispectrum to the
covariance matrix \cite{Lewis_2011}. We shall ignore this contribution
for simplicity, as our primary goal here is to investigate how
non-linear RS effect changes the signal-to-noise ratio relative to the
linear ISW effect. As a result, our signal-to-noise ratio is overestimated
by 10\% at $l_{\rm max}=1500$ and by $\gtrsim 40$\% at $l_{\rm max}\gtrsim 2000$.
As all the previous work except for \cite{Lewis_2011} has also ignored 
this contribution to the covariance matrix, our results for the signal-to-noise
ratio can be compared directly with those from the previous work.

In Figure~\ref{sum_plots_PT_MA99_HALOFIT}, we show $S/N$ as a function
of the maximum multipole, $l_{\rm max}$. We find that all non-linear
models as well as the linear model give similar results. This confirms
our earlier observation (see Sec.~\ref{LRS-CMB_bispectrum}) that the
lensing-RS bispectrum peaks in the squeezed limit where the smallest
multipole corresponds to $l$ of $Q(l)$, and thus most of the
signal-to-noise comes from the region where $Q(l)$ is well approximated
by the linear lensing-ISW cross-correlation power spectrum. Our $S/N$
estimate agrees well with that from the linear calculation of
\cite{Smith_Zaldarriaga_2006}. 

Note that our $S/N$ for the linear model is about a factor of 2.7
smaller than that of \cite{Mangilli_Verde_2009}.\footnote{For this comparison, 
we use Eq.~(\ref{RS_bispectrum}) instead of Eq.~(\ref{RS_bispectrum2}) because they have used
the unlensed $C_l$ when calculating the lensing-RS bispectrum.} Their Eq.~(15) suggests
that they have not restricted the sum to $l_1 \leq l_2 \leq l_3 $, which
results in an overestimation of $(S/N)^2$ by a factor of 6, i.e., a
factor 2.4 in $S/N$, which is enough to explain the difference. 

Our $S/N$ for the non-linear model using MA99 is an order of magnitude
smaller than that of \cite{Giovi_2005}.$^5$ This is probably due to a
combination of their not restricting the sum to $l_1 \leq l_2 \leq l_3 $
and their overestimating the bispectrum with an incorrect implementation
of MA99 (see Sec.~\ref{sec:ma99}). 

\subsection{$\chi^2$ difference between linear and non-linear models}
Can we detect differences between the linear and non-linear models? In
order to answer this question, we calculate the $\chi^2$ differences
between the linear and non-linear models from \cite{Mangilli_Verde_2009}
\begin{equation}
\label{chi_squared_eq}
\chi_{\mathrm{X-Y}}^2 \equiv \sum_{2 \leq l_1 \leq l_2 \leq l_3 < l_{\mathrm{max}}} { 
{\frac{ \left( B^{\mathrm{X}}_{l_1 l_2 l_3} - B^{\mathrm{Y}}_{l_1 l_2 l_3}\right)^2  }{\Delta_{l_1l_2l_3}C_{l_1}C_{l_2}C_{l_3} }},
}
\end{equation}
where $\mathrm{X}$ and $\mathrm{Y}$ denote the names of models under
consideration. For example, when we study the $\chi^2$ difference
between 3PT and the linear model, $X={\rm 3PT}$ and $Y={\rm LIN}$. 

%%%%%%%%%%%%%%%%%%%%%%%%%%%%%%%%%%%%%%%%%%%%%%%%%%%%%%%%%%%%%%%%%% 
\begin{figure}[t]
\begin{center}
\includegraphics[width=0.49\textwidth]{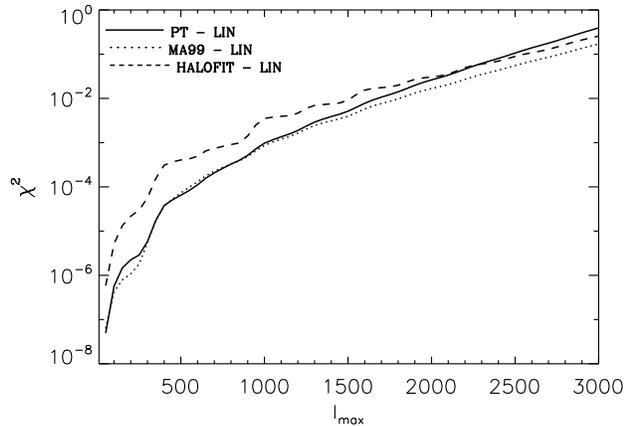}
\caption[]
{ The $\chi^2$ differences between the linear model and various non-linear
 models: 3PT (solid line), MA99 (dotted line), and HALOFIT (dashed
 line). 
}
\label{diff_plots_PT_MA99_HALOFIT_lin}
\end{center}
\end{figure}
%%%%%%%%%%%%%%%%%%%%%%%%%%%%%%%%%%%%%%%%%%%%%%%%%%%%%%%%%%%%%%%%%%%%

In Figure~\ref{diff_plots_PT_MA99_HALOFIT_lin}, we show the $\chi^2$
differences between the linear model and various non-linear models:
$\chi^2_{X-\rm LIN}$ for $X={\rm 3PT}$, MA99, and HALOFIT. We find that, for all non-linear models, the
$\chi^2$ differences are much smaller than unity, indicating that the
differences are too small to detect. We find similar values of $\chi^2$
differences among non-linear models. 

Our results do not agree with those of \cite{Mangilli_Verde_2009}, who
find $\chi^2_{X-\rm LIN}$ of order unity for $X={\rm HALOFIT}$. We suspect
that this is potentially due to (i) their not restricting the sum to $l_1
\leq l_2 \leq l_3$ (which would account for a factor of 6), and (ii)
a numerical accuracy of their evaluation
of $\partial P_\Psi(k,z)/\partial z$. As noted in
Sec.~\ref{sec:derivative}, a simple derivative such as $\partial
P_\Psi(k,z)/\partial z =[P_\Psi(k,z+\delta z)-P_\Psi(k,z)]/\delta z$ can
result in an inaccurate result, and a better method such as the cubic
spline interpolation is needed for correctly calculating this
derivative. We have confirmed this by using the above simple derivative,
finding that the results can vary significantly if such simpler numerical
derivatives are used.

\section{Contamination of the local-form primordial non-Gaussianity}
\label{fNL}

While the lensing-ISW and lensing-RS bispectra are useful for studying
the nature of dark energy, they are also important because they yield
the largest known contamination of the so-called ``local form''
primordial bispectrum
\cite{Smith_Zaldarriaga_2006,Serra_Cooray_2008,Hanson_2009}. The
local-form bispectrum is particularly important, as a significant
detection of the {\it primordial} bispectrum of this form would rule out
all single-field inflation models regardless of the details of models
\cite{Creminelli_Zaldarriaga_2004}. 

%%%%%%%%%%%%%%%%%%%%%%%%%%%%%%%%%%%%%%%%%%%%%%%%%%%%%%%%%%%%%%%%%% 
\begin{figure}[t]
\begin{center}
\includegraphics[width=0.49\textwidth]{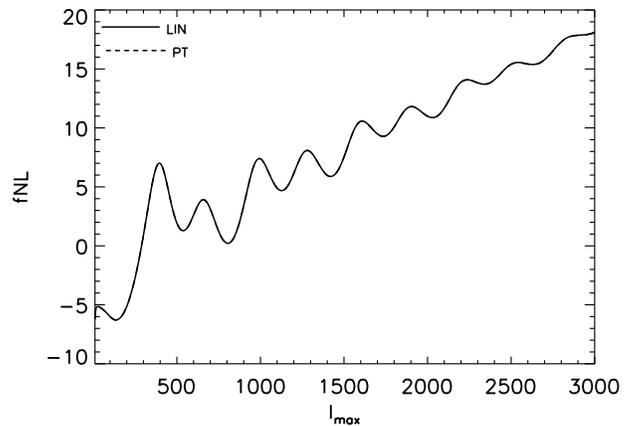}
\caption[]
{ The contamination of $f_{\rm NL}$ due to the lensing-ISW (linear
 effect only; solid line) and the lensing-RS (3PT;
 dashed line). Non-linearity does not affect the contamination of
 $f_{\rm NL}$. The solid and dashed lines are indistinguishable.
}
\label{fig_fNL}
\end{center}
\end{figure}
%%%%%%%%%%%%%%%%%%%%%%%%%%%%%%%%%%%%%%%%%%%%%%%%%%%%%%%%%%%%%%%%%%%%

The contamination of the local-form primordial bispectrum, parametrized
by the $f_{\rm NL}$ parameter, has been computed for the lensing-ISW
bispectrum. How would non-linearity (lensing-RS) affect $f_{\rm NL}$? To
answer this question, we calculate the ``bias in $f_{\rm NL}$,'' i.e., a
value of $f_{\rm NL}$ which would be found if we fit the lensing-RS
bispectrum to the local-form primordial bispectrum template:
\begin{equation}
 \delta f_{\rm NL} = \frac{\sum_{l_1\le l_2\le l_3}\frac{B_{l_1l_2l_3}^{\rm
  prim}B_{l_ll_2l_3}^{\rm lens-RS}}{\Delta_{l_1l_2l_3}C_{l_1}C_{l_2}C_{l_3}}}{\sum_{l_1\le l_2\le l_3}\frac{(B_{l_1l_2l_3}^{\rm  prim})^2}{\Delta_{l_1l_2l_3}C_{l_1}C_{l_2}C_{l_3}}},
\end{equation}
where $B_{l_1l_2l_3}^{\rm prim}$ is the local-form primordial bispectrum
given in \cite{Komatsu_Spergel_2001}.

In Figure~\ref{fig_fNL}, we show the contamination of $f_{\rm NL}$ for the
lensing-ISW (solid) and the lensing-RS (computed with 3PT; dashed)
bispectra. We find that 
they give similar results, and thus non-linearity does not affect the
contamination of $f_{\rm NL}$. The values of $\delta f_{\rm NL}(l_{\rm
max})$ that we find agree well with those from the linear calculation of
\cite{Hanson_2009}.  

%%%%%%%%%%%%%%%%%%%%%%%%%%%%%%%%%%%%%%%%%%%%%%%%%%%%%%%%%%%%%%%%%% 
%%%%%%%%%%%%%%%%%%%%%%%%%%%%%%%%%%%%%%%%%%%%%%%%%%%%%%%%%%%%%%%%%% 
%%%%%%%%%%%%%%%%%%%%%%%%%%%%%%%%%%%%%%%%%%%%%%%%%%%%%%%%%%%%%%%%%% 
\section{Conclusions}
\label{conclusions}
%%%%%%%%%%%%%%%%%%%%%%%%%%%%%%%%%%%%%%%%%%%%%%%%%%%%%%%%%%%%%%%%%% 
%%%%%%%%%%%%%%%%%%%%%%%%%%%%%%%%%%%%%%%%%%%%%%%%%%%%%%%%%%%%%%%%%% 
%%%%%%%%%%%%%%%%%%%%%%%%%%%%%%%%%%%%%%%%%%%%%%%%%%%%%%%%%%%%%%%%%% 

The basic findings of this paper are simple: while non-linear matter
clustering modifies the shape of the lensing-ISW bispectrum, differences
between the linear prediction and non-linear predictions as well as
differences among non-linear predictions are too small
to detect. Non-linearity does not affect the contamination of the
local-form primordial bispectrum. This is because the lensing-ISW
bispectrum peaks in the squeezed configuration in which the smallest
multipole corresponds to the multipole of the lensing-ISW
cross-correlation power spectrum, where the linear approximation is
valid. Therefore, the linear calculation would be practically sufficient
when interpreting  the CMB data such as those from Planck. 

Nevertheless, if one wishes to improve upon the
linear calculation, one should probably use the 3PT, as it offers a greater
flexibility in terms of cosmological models for which the calculations
are valid, as well as a straightforward computation of $\partial
P_\Psi(k,z)/\partial z$.

Our results do not agree with the previous work studying the lensing-RS
bispectrum
\cite{Verde_Spergel_2002,Baccigalupi_2003,*Giovi_2005,Mangilli_Verde_2009}
which found much greater effects of non-linear clustering on the
lensing-RS bispectrum. We suspect that the discrepancy is due to a
combination of 
an incorrect implementation of MA99, an inaccurate numerical evaluation of
$\partial P_\Psi(k,z)/\partial z$, and/or their not restricting the sum
in $(S/N)^2$ to $l_1\le l_2\le l_3$. As a result, the expected
signal-to-noise ratio of the lensing-RS 
bispectrum for a cosmic-variance-limited experiment is about 5 for
$l_{\rm max}=1500$, which is smaller than the previous estimates
\cite{Baccigalupi_2003,*Giovi_2005,Mangilli_Verde_2009}, but agrees well
with the estimates based on the linear calculation
\cite{Smith_Zaldarriaga_2006,Serra_Cooray_2008,Hanson_2009}. 

We would like to thank J. Weller and D. Spergel for discussions. We
would like to thank A. Lewis for pointing out that we should use the
lensed $C_l$ in Eq.~(\ref{RS_bispectrum2}) as well as the importance of the
non-Gaussian contribution to the covariance matrix given in
Eq.~(\ref{Signal_to_Noisse_komat}). 
This work is supported in part by NSF grant PHY-0758153.  

%\bibliography{literature}

%merlin.mbs apsrev4-1.bst 2010-07-25 4.21a (PWD, AO, DPC) hacked
%Control: key (0)
%Control: author (8) initials jnrlst
%Control: editor formatted (1) identically to author
%Control: production of article title (-1) disabled
%Control: page (0) single
%Control: year (1) truncated
%Control: production of eprint (0) enabled
%

\end{document}